\documentclass[%
 reprint,
 amsmath,amssymb,
 aps,
 pre,
]{revtex4-1}

\usepackage{graphicx}
\usepackage{color}
\usepackage{dcolumn}
\usepackage{bm}

\begin{document}


\title{Periodic potential can enormously boost \\free particle transport induced by active fluctuations}

\author{K. Bia{\l}as}
\affiliation{Institute of Physics, University of Silesia, 41-500 Chorz{\'o}w, Poland}
\author{J. {\L}uczka}
\affiliation{Institute of Physics, University of Silesia, 41-500 Chorz{\'o}w, Poland}
\author{J. Spiechowicz}
\email{jakub.spiechowicz@us.edu.pl}
\affiliation{Institute of Physics, University of Silesia, 41-500 Chorz{\'o}w, Poland}

\begin{abstract}
Active fluctuations are detected in a growing number of systems due to self-propulsion mechanisms or collisions with active environment. They drive the system far from equilibrium and can induce phenomena which at equilibrium states are forbidden by e.g. fluctuation-dissipation relations and detailed balance symmetry. Understanding of their role in living matter is emerging as a challenge for physics. Here we demonstrate a paradoxical effect in which a free particle transport induced by \emph{active fluctuations} can be boosted by many orders of magnitude when the particle is additionally subjected to a periodic potential.
In contrast, within the realm of only \emph{thermal fluctuations} the velocity of a free particle exposed to a bias is reduced when the periodic potential is switched on. The presented mechanism is significant for understanding nonequilibrium environments such as living cells where it can explain from a fundamental point of view why spatially periodic structures known as microtubules are necessary to generate impressively effective intracellular transport. Our findings can be readily corroborated experimentally e.g. in a setup comprising a colloidal particle in an optically generated periodic potential.
\end{abstract}

\maketitle


\section{Introduction}
Microscopic systems are inherently immersed in a sea of thermal fluctuations that may strongly influence their properties or even induce an entirely new phenomenology. Celebrated examples include stochastic \cite{benzi1981, gammaitoni1998} or coherence \cite{pikovsky1997,lindner2004} resonance, anomalous diffusion \cite{bouchaud1990,metzler2014,spiechowicz2019njp}, ratchet effects \cite{hanggi2009,reichhardt2017,spiechowicz2016scirep}, negative mobility \cite{machura2007,speer2007,nagel2008,slapik2019} or thermal noise induced dynamical localization \cite{spiechowicz2017scirep,spiechowicz2019chaos}, to name only a few. Nevertheless impact of thermal equilibrium fluctuations is limited by fundamental laws of nature such as the fluctuation-dissipation theorem \cite{kubo1966,marconi2008} or the detailed balance symmetry \cite{cates2012,gnesotto2018}.

These restrictions are no longer true for a nonequilibrium environment which keeps the system permanently out of thermal equilibrium even in the absence of external perturbations. A prototypical example are active fluctuations manifesting themselves either in (i) active matter harvesting energy from the environment into a self-propulsion drive \cite{ramaswamy2010, romanczuk2012, cates2012, marchetti2013, bechinger2016, reichhardt2017} or (ii) active bath such as a suspension of active colloids or swimming microorganisms (e.g. bacteria \emph{Escherichia coli}) pushing around a passive system \cite{bechinger2016, maggi2014, kanazawa2015, maggi2017, dabelow2019}. Their relevance in biological systems is emerging as a \emph{hot topic} which is in the focus of researchers across all branches of natural science \cite{kanazawa2020}. For instance, the latest breaking experimental results make it clear that active fluctuations in living cells, generated by using energy derived from metabolic activities, are not just noise but are utilized to promote various physiological processes. In particular, biological motors such as kinesin or dynein benefit from active fluctuations and enhance their directional movement \cite{ezber2020, ariga2021}.

Within the realm of \emph{thermal equilibrium fluctuations} the directed velocity of a free particle exposed to a bias is usually reduced or at best conserved when an additional periodic potential is switched on \cite{risken}. Similarly, the diffusion coefficient $D_0$ of the free Brownian particle subjected to a periodic force is reduced to its effective diffusion constant $D<D_0$ \cite{lifson,festa}. In this work we demonstrate an opposite striking case: the particle can harness \emph{active nonequilibrium fluctuations} to exploit an unbiased periodic potential and enhance its directed velocity much larger than for free transport. We identify its origin in smart nature of active fluctuations which unlike thermal ones are not constrained by the equilibrium state and therefore may optimize themselves to make use of a periodic potential.

Our results are significant for understanding nonequilibrium environments and addressing important theoretical questions concerning thermodynamics of active systems such as molecular motors inside living cells. They can be corroborated experimentally in a setup comprising a colloidal particle in an optically generated periodic potential \cite{park2020,paneru2021} or tested for real biological motors \cite{ezber2020, ariga2021}. Moreover, they may open new avenues for designing ultrafast and efficient micro and nanoscale machines.

The paper is organized as follows. In the next section we discuss details of the studied model. In Sec. III we present the results. Firstly, we show a paradoxical effect in which the periodic structure not only does not hinder the directed velocity of the particle, but on the contrary, it is involved in inducing the giant transport which can be orders of magnitude greater than the velocity of the free particle. Secondly, we outline the mechanism of this phenomenon. Thirdly, we discuss the role of breaking of the periodic potential and active fluctuations symmetry. The last section provides the summary and conclusions. In Appendix A we introduce the dimensionless units whereas in Appendix B we detail on the parametrization of active fluctuations amplitude distribution. Finally, in the last Appendix we discuss exemplary realizations of active fluctuations.

\section{Model}
We start our considerations with overdamped motion of a free particle subjected to both active nonequilibrium fluctuations $\eta(t)$ and thermal noise $\xi(t)$,   
\begin{equation}
	\dot{x}=\eta(t) + \sqrt{2 D_T}\,\xi(t),
    \label{eq0}
\end{equation}
where the dot denotes differentiation with respect to the time $t$. Thermal fluctuations $\xi(t)$ of temperature $D_T$ are modeled by Gaussian white noise of zero mean $\langle \xi(t) \rangle = 0$ and the correlation function $\langle \xi(t)\xi(s) \rangle = \delta(t-s)$. Details of the scaling procedure are presented in Appendix A. We assume that $\langle \eta(t)\rangle = v_0 \ne 0$ and in consequence active noise $\eta(t)$ induces the directed transport of the Brownian particle.   Then one finds $\langle \dot{x}\rangle = v_0$. The essence of the proposed strategy for enhancement of the directed transport  is to impose an unbiased periodic potential $U(x) = U(x + 2\pi)$ to the system (\ref{eq0}), i.e., 
\begin{equation}
	\dot{x}=-\varepsilon U'(x) + \eta(t) + \sqrt{2 D_T}\,\xi(t),
    \label{eq1}
\end{equation}
where $\varepsilon$ is a half of the potential barrier height and the prime denotes differentiation with respect to the position $x$ of the particle. First, we consider the simplest symmetric periodic potential
\begin{equation}
	U(x) = \sin{x}
\end{equation}
in order to avoid the apparent ratchet effect. 
Ratchet potentials which break the spatial symmetry should additionally enhance the free particle transport and will be analyzed subsequently. 

We pose a question for which model of active fluctuations $\eta(t)$ the directed transport can be enormously enhanced, i.e. $\langle \dot x \rangle \gg v_0$. As a candidate for $\eta(t)$ we choose Poisson white noise \cite{hanggi1978,spiechowicz2014pre,bialas2020} which is a random sequence of $\delta$-shaped pulses with random amplitudes $z_i$, 
\begin{equation}
    \eta(t)=\sum_{i=1}^{n(t)}z_i\delta(t-t_i),
\end{equation}
where $t_i$ are the arrival times of Poisson counting process $n(t)$, namely, the probability for emergence of $k$ impulses in the time interval $[0,t]$ is  $Pr\{n(t)=k\}=(\lambda t)^k \exp{(-\lambda t)}/k!$ \cite{feller1970}. The parameter $\lambda$ is the mean number of $\delta$-spikes per unit time (the firing rate of the Poisson process). The amplitudes $\{z_i\}$ of $\delta$-kicks are statistically independent random variables sampled from the common probability distribution $\rho(z)$. The process $\eta(t)$ presents white noise of a finite mean and a covariance given by
\begin{equation}
	\label{pwsnmom}
	\langle \eta(t) \rangle = \lambda \langle z_i \rangle, \quad \langle \eta(t)\eta(s) \rangle  - \langle \eta(t) \rangle \langle \eta(s) \rangle = 2D_P\delta(t-s),
\end{equation}
where $\langle z_i \rangle$ is an average over the amplitude distribution $\rho(z)$ and noise intensity is $D_P = \lambda \langle z_i^2 \rangle/2$. It is statistically symmetric if the density  $\rho(z)$ is symmetric, i.e. when $\rho(z) = \rho(-z)$. In such a case $\langle \eta(t) \rangle = 0$. However, we have to consider biased Poissonian noise for which $\langle \eta(t) \rangle = v_0 \ne  0$. Because there are infinitely many different statistics of the amplitudes $\{z_i\}$ for which $\langle \eta(t) \rangle = v_0$, we pick one of the most general forms of $\rho(z)$, namely, the skew-normal distribution \cite{hagan,azz,rijal2022,bailey2021} which allows to take into account both positive and negative amplitudes $z_i$ as well as asymmetry in the distribution $\rho(z)$, c.f. Appendix B. It is characterized by three independent parameters: the mean value $\langle z_i \rangle = \zeta$, variance $\sigma^2$ and asymmetry (skewness) $\chi$ \cite{generacja,generacja2}. Active fluctuations $\eta(t)$ defined in such a way  are represented by  white noise of average  $\langle \eta(t) \rangle = \lambda \zeta$. We also assume that thermal and active fluctuations are uncorrelated, $\langle \xi(t) \eta(s) \rangle = \langle \xi(t) \rangle \langle \eta(s) \rangle = 0$. Noise $\eta(t)$ is applied in Ref. \cite{lee2022} and it can serve as a model for stochastic release of energy in chemical reactions such as ATP hydrolysis or random collisions with complex and crowded environments. In this sense our model is appropriate for both an active particle self-propelling itself inside a passive medium or a passive system immersed in active bath formed as a suspension of active particles \cite{dabelow2019}.
\begin{figure}[t]
	\centering
	\includegraphics[width=0.9\linewidth]{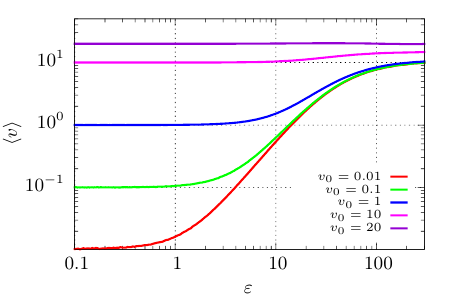}
	\caption{The average velocity $\langle v \rangle$ of the Brownian particle dwelling in the  symmetric potential $U(x) = \sin{x}$ and driven by active fluctuations $\eta(t)$ vs the potential barrier height $\varepsilon$ for selected values of $\langle \eta(t) \rangle = v_0$. Parameters read: the spiking rate $\lambda = 30$, variance and skewness of the amplitude distribution $\rho(z)$ is $\sigma^2 = 3.1$ and $\chi = 0.99$, respectively. Temperature is set to $D_T = 0.01$.}
	\label{fig1}
\end{figure}

The main quantity of interest for the study of transport properties is the directed velocity defined as \mbox{$\langle v \rangle =\lim_{t\to\infty}\langle x(t) \rangle/t$}, where $\langle \cdot \rangle$ indicates the ensemble average. If $\eta(t) = 0$ in Eq. (\ref{eq1}) then $\langle v \rangle = 0$. In the presence of active fluctuations with $\langle \eta(t) \rangle = v_0$ the free Brownian particle is transported with the average velocity $\langle v \rangle = v_0$. If under the identical experimental conditions the periodic potential $U(x)$ is turned on, one expects that the directed velocity will be notably reduced $\langle v \rangle < v_0$, in particular for $U(x)$ with large barrier height $\varepsilon$ \cite{risken}.
 
\section{Results}
Unfortunately, the Fokker-Planck-Kolmogorov-Feller integro-differential equation corresponding to Eq. (\ref{eq1}) cannot be solved analytically in a closed form \cite{hanggi1978}. Therefore we resort to precise numerical simulations done by harvesting the GPU supercomputer using the CUDA enviroment \cite{spiechowicz2015cpc}. The ensemble averaging was performed over Gaussian $\xi(t)$ and Poissonian $\eta(t)$ noise realizations as well as over the initial conditions $x(0)$ distributed uniformly over the spatial period $[0, 2\pi]$ of the potential $U(x)$.
\begin{figure}[t]
    \centering
    \includegraphics[width=0.9\linewidth]{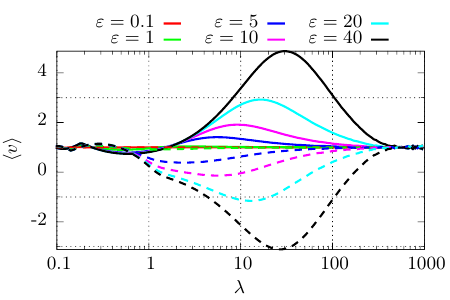}
    \caption{The average velocity $\langle v \rangle$ versus the spiking rate $\lambda$ of active fluctuations $\eta(t)$ for selected values of the potential barrier height $\varepsilon$ and fixed $\langle \eta(t) \rangle = v_0 = 1$. Solid lines correspond to the asymmetry $\chi=0.99$, whereas dashed ones to $\chi=-0.99$. Other parameters are the variance $\sigma^2=3.1$ and temperature $D_T = 0.01$.}
    \label{fig2}
\end{figure}
\begin{figure}[t]
    \centering
    \includegraphics[width=0.9\linewidth]{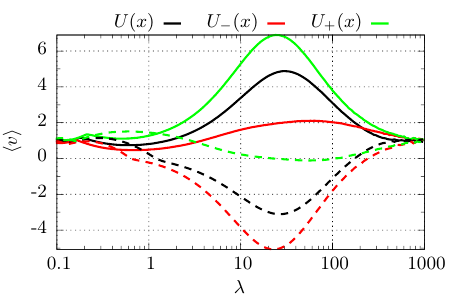}
    \caption{Illustration of the role of potential symmetry on enhancement of free particle  transport. The potential barrier is \mbox{$\varepsilon = 40$}. The average bias $\langle \eta(t) \rangle = v_0 = 1$. Solid lines correspond to the asymmetry $\chi=0.99$, whereas dashed ones to $\chi=-0.99$. Other parameters read: the variance $\sigma^2 = 3.1$ and temperature is fixed to $D_T = 0.01$.}
    \label{fig5}
\end{figure} 

In Fig. \ref{fig1} we present an  effect that contradicts common intuition in which the periodic structure not only does not hinder the directed velocity of the particle, but on the contrary, it is involved in inducing the \emph{giant transport} which can be several orders of magnitude greater than the velocity of the free particle, i.e., $\langle v \rangle \gg v_0$.
We note that when the barrier $\varepsilon$ is increased the average velocity $\langle v \rangle$ can be remarkable enhanced. E.g. for $\langle \eta(t) \rangle = 0.01$ and $\varepsilon = 100$ the particle velocity reads $\langle v \rangle \approx 10 \gg v_0=0.01$, i.e. three orders of magnitude greater than for the free particle driven by the same active fluctuations $\eta(t)$. This giant transport behavior $\langle v \rangle \gg v_0$ is clearly induced by the additional presence of the symmetric periodic potential $U(x) = \sin{x}$ with deep wells. In Fig. \ref{fig1} we show that amplification of $\langle v \rangle$ depends on $v_0$. For small $v_0 \ll 1$ the velocity $\langle v \rangle$ can be enormously boosted when the system is additionally subjected to a periodic potential $U(x)$. In such a case $\langle v \rangle(\varepsilon)$ increases rapidly when $\varepsilon$ grows. On the other hand, if the free particle transport is already large,  $v_0 \gg 1$, then there is no gain by placing the system in a periodic potential.

In many nonequilibrium systems, especially in living cells, there are no substantial systematic gradients or forces that could induce large transport. Instead one typically encounter unbiased or weakly-biased fluctuations of both thermal and active (non-thermal) origin. Therefore the presented mechanism might be crucial for deep understanding of their physics. In Sec. III A we show that this non-trivial effect is heavily influenced by the statistics $\rho(z)$ of amplitudes $z_i$ of fluctuations $\eta(t)$. In particular, it is not observed when $z_i$  are distributed according to the frequently applied  and even more asymmetric exponential probability density.  

To explain the mechanism of this phenomenon we first fix the mean of active fluctuations $\langle \eta(t) \rangle = 1$ and study how the velocity $\langle v \rangle$ depends on the spiking rate $\lambda$. This characteristics is depicted in \mbox{Fig. \ref{fig2}}. If $\lambda \to 0$ or \mbox{$\lambda \to \infty$} then  $\langle v \rangle \to v_0$. Since the bias of nonequilibrium noise is fixed $\langle \eta(t) \rangle =  \lambda \zeta = 1$  the average amplitude $\zeta$ of $\delta$-spikes goes to infinity when $\lambda \to 0$ and therefore the potential $U(x)$ becomes negligible. On the other hand, when  $\lambda \to \infty$ one stacks the infinite number of $\delta$-kicks with vanishing mean but the fixed variance $\sigma^2 = 3.1$ and the resultant movement mimics the free particle. The most striking feature of this panel is that for growing potential barrier $\varepsilon$ there is the optimal spiking rate $\lambda_{max}(\varepsilon)$ for which the velocity is maximal and significantly greater than the free particle velocity $\langle v \rangle > v_0$. We stress that for smaller values of $\langle \eta(t) \rangle$ one observes the giant enhancement $\langle v \rangle \gg v_0$, c.f. Fig. \ref{fig1}. The effect of $\varepsilon$-increase is two-fold. Firstly, it shifts the optimal $\lambda_{max}(\varepsilon)$ towards the larger values and secondly, the magnitude of the velocity enhancement grows at $\lambda=\lambda_{max}(\varepsilon)$.
\begin{figure}[t]
	\centering
	\includegraphics[width=0.9\linewidth]{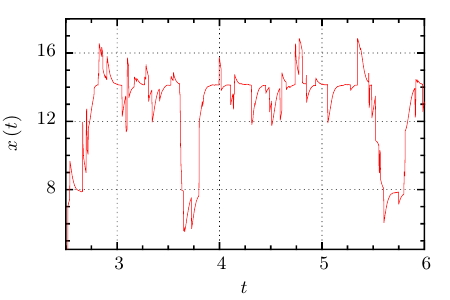}
	\caption{Exemplary Brownian particle trajectory is shown for the potential barrier $\varepsilon = 40$. Other parameters read:  the spiking rate $\lambda = 30$, the average bias $\langle \eta(t) \rangle = v_0 = 1$, the variance $\sigma^2 = 3.1$ and asymmetry $\chi = 0.99$ of the amplitude distribution $\rho(z)$. Temperature is fixed at $D_T = 0.01$.}
	\label{fig3}
\end{figure} 

In Fig. \ref{fig2} one can observe also the impact of the asymmetry $\chi$ of the amplitude distribution $\rho(z)$, see Appendix B. At first glance it may seem that the transport direction is completely determined by the sign of mean $\langle \eta(t) \rangle = \lambda \zeta$ of active fluctuations which is controlled by the average value $\langle z_i \rangle = \zeta$ of $\delta$-spikes distributed according to the probability density $\rho(z)$. However, it is not true in general. In particular, even though the statistical bias reads $\langle \eta(t) \rangle = v_0 = 1$, when $\chi$ is reversed from the positive $\chi = 0.99$ to the negative value $\chi = -0.99$ the transport direction is also inverted, i.e. $\langle v \rangle < 0$. Therefore we conclude that in the studied case the orientation of long tail in $\rho(z)$ is responsible for pointing the direction of particle movement, c.f. Appendix B. When the asymmetry is negative the transport enhancement over the velocity of free particle $\langle v \rangle > v_0$ is still observed, however, its magnitude is a little bit smaller than for the situation when $\chi$ is positive. 

In Fig. \ref{fig3} we present the exemplary Brownian particle trajectory in the studied parameter regime with \mbox{$\langle \eta(t) \rangle = 1$} and the large potential barrier $\varepsilon = 40$. Its careful inspection suggests that the occurring transport process may be decomposed onto two contributions. The first one is associated with the arrival of $\delta$-kick because of which the particle overcomes the potential barrier. In principle in this way it can be carried over its many spatial periods. However, the second one which is clearly missing for the free particle, is related to its relaxation towards the nearest potential minimum as it is e.g. for $t \approx 2.6$ or $t \approx 5.6$. This contribution is at the root of the giant transport effect. When the mean value $\langle \eta(t) \rangle = v_0$ of active fluctuations is large then regardless of the potential barrier $\varepsilon$ magnitude the particle velocity attains its free transport value $\langle v \rangle = v_0$, c.f. Fig \ref{fig1}. It is due to the fact that in such a limit the contribution coming from the relaxation is negligible. However, when $\langle \eta(t) \rangle = v_0$ is small, see e.g. $\langle \eta(t) \rangle = v_0 = 0.01$ in Fig. \ref{fig1}, the sliding towards the potential minima plays an essential role and the giant transport occurs when the potential barrier $\varepsilon$ is increased. 
\begin{figure}[t]
    \centering
    \includegraphics[width=0.9\linewidth]{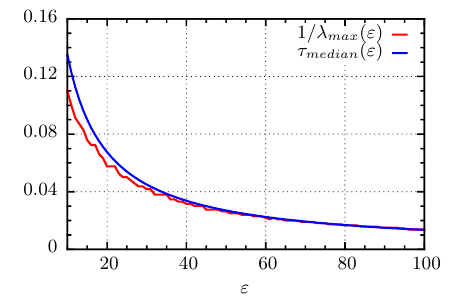}
    \caption{The characteristic time $1/\lambda_{max}(\varepsilon)$ between two successive $\delta$-kicks of active fluctuations $\eta(t)$ and the estimated median time $\tau_{median}(\varepsilon)$ of particle relaxation towards the potential $U(x)$ minimum both as a function of the barrier height $\varepsilon$. Other parameters read: the average value of active fluctuations $\langle \eta(t) \rangle = v_0 = 1$, variance $\sigma^2=3.1$ and skewness $\chi = 0.99$ of the amplitude distribution $\rho(z)$, temperature $D_T = 0.01$.}
    \label{fig4}
\end{figure}
\begin{figure}[t]
    \centering
    \includegraphics[width=1.0\linewidth]{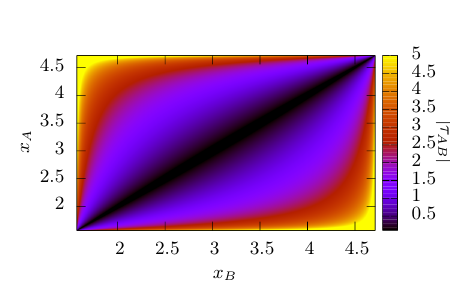}
    \caption{Absolute value of the relaxation time $|\tau_{AB}|$ from the position $x_A$ to $x_B$ calculated according to Eq. (\ref{eq_S_sol}).}
    \label{fig:S4}
\end{figure}

To illustrate this fact we ask about the physical interpretation of the optimal spiking rate $\lambda_{max}(\varepsilon)$ for which the rescaled particle velocity $\langle v \rangle/v_0$ assumes its maximal value when $\langle \eta(t) \rangle = v_0$ is fixed. In Fig. \ref{fig4} we present inverse of this characteristics $1/\lambda_{max}(\varepsilon)$ and the estimated median time $\tau_{median}(\varepsilon)$ of particle relaxation towards the potential minimum both versus the barrier height $\varepsilon$. In absence of all fluctuations the latter relaxation for the potential $U(x)=\sin(x)$ with the barrier height $\varepsilon$ is determined by the equation 
\begin{equation}
    \dot{x}=-\varepsilon U'(x) = - \varepsilon \cos{x}.
    \label{eq_S_diff}
\end{equation}
The time $\tau_{AB}$ the particle needs to move from the point $x_A$ to $x_B$ reads
\begin{equation}
\tau_{AB}=-\frac{1}{\varepsilon}\int_{x_A}^{x_B} \frac{dx}{\cos{x}}=-\frac{1}{2\varepsilon}\ln \left| \frac{1+\sin{x}}{1-\sin{x}}\right|\Biggr|_{x_A}^{x_B}.
    \label{eq_S_sol}
\end{equation}
After the arrival of $\delta$-spike the particle can land at any random position $x_A$ and then during the time interval $\tau_{AB}$ it is relaxing towards the nearest potential minimum. This process ends at the random position $x_B$ where another $\delta$-spike of active nonequilibrium noise $\eta(t)$ emerges. As the potential is periodic we restrict ourselves to the interval $x_A,x_B \in\left(\frac{\pi}{2},\frac{3\pi}{2}\right)$. Note that both the minimum and maximum are excluded because the time required to leave the maximum or reach the minimum is infinite and Eq. (\ref{eq_S_sol}) does not converge for these values. In Fig. \ref{fig:S4} we present the relaxation time $|\tau_{AB}|$ for every pair of the starting and ending points taken from the considered interval $x_A,x_B \in\left(\frac{\pi}{2},\frac{3\pi}{2}\right)$. From this characteristic the mean $\tau_{mean}(\varepsilon) \approx 1.72/\varepsilon$ and median $\tau_{median}(\varepsilon) \approx1.35/\varepsilon$ of the relaxation time is evaluated.  

In Fig. \ref{fig4} one can clearly see that the initial discrepancy between these two characteristic time scales quickly dies out as $\varepsilon$ is increased and they become equivalent when the giant transport is detected. Such resonance-like behaviour explains the mechanism of this counterintuitive effect. It means that the enhancement of particle velocity over free transport is maximal when the average time  $1/\lambda_{max}(\varepsilon)$ between two successive $\delta$-kicks of active fluctuations matches the interval $\tau_{median}(\varepsilon)$ needed for the particle to exploit the process of relaxation towards the potential minimum. The resulting motion is synchronized: the particle is $\delta$-kicked and fall on one of the potential  slopes, in the next time interval statistically there are no other $\delta$-spikes and it relaxes to a neighbouring minimum of the potential  and this scenario repeats over and over again. We now demonstrate that this mechanism is non-trivial and does not emerge e.g. for the exponential distribution  $\rho(z)$ of amplitudes $z_i$ which is even more asymmetric than the skew-normal distribution and arises in numerous different contexts.

\subsection{Role  of ratchet potentials}
The next factor which can additionally boost the particle velocity is related to the potential symmetry. This issue is presented in Fig. \ref{fig5} for three forms of the periodic potential:  $U(x)$ is symmetric whereas both $U_+(x)$ and $U_-(x)$ depict the asymmetric ratchet \cite{luczka1995} and read
\begin{subequations}
\begin{align}
U_{+}(x) &= 0.908 [\sin{x} + 0.25\sin{(2x)}], \\ 
U_{-}(x) &= 0.908 [\sin{x} - 0.25\sin{(2x)}]
\end{align}
\end{subequations}
The prefactor $0.908$ in $U_{+}(x)$ and $U_{-}(x)$ ensures that they possess the same barrier height as $U(x)$. 
The most important conclusion is that the giant increase of free transport depends on the particular realization of periodic substrates and for an appropriate choice it can be even further amplified. E.g. for the ratchet $U_+(x)$ the velocity $\langle v \rangle$ is noticeably larger than for the symmetric potential $U(x)$. By comparing these three potentials one can notice that the distance between their maxima $x_{max}$ and subsequent minima $x_{min} > x_{max}$ is greater for the ratchet $U_+(x)$ than for $U(x)$. Conversely, for $U_-(x)$ it is smaller than for $U(x)$. It is more likely that active fluctuations $\eta(t)$ will kick out the particle on the longer slope of the potential. Therefore if $\chi = 0.99$ and the transport occurs in the positive direction, the velocity $\langle v \rangle$ for $U(x)$ is greater than for $U_-(x)$ and smaller than for $U_+(x)$. As we just explained when $\chi = -0.99$ the particle moves in the negative direction and the situation is reversed.
\begin{figure}[t]
	\centering
	\includegraphics[width=0.9\linewidth]{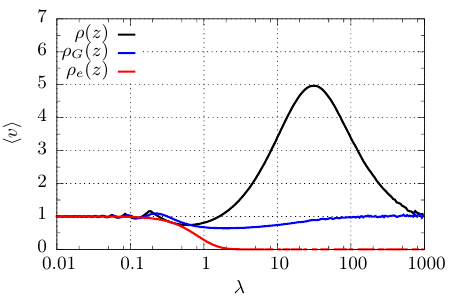}\\
	\includegraphics[width=0.9\linewidth]{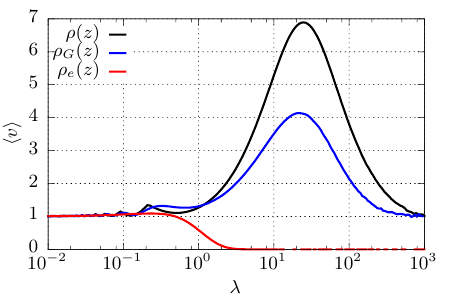}
	\caption{The average velocity $\langle v \rangle$ versus the spiking rate $\lambda$ of active fluctuations $\eta(t)$ depicted for three amplitude distributions and fixed $\langle \eta(t) \rangle = v_0 = \lambda \zeta = 1$, i.e. $\zeta = 1/\lambda$. $\rho(z)$ corresponds to the skew-normal statistics with $\sigma^2 = 3.1$ and $\chi = 0.99$; $\rho_G(z)$ indicates the Gaussian density with $\sigma^2 =3.1$; $\rho_e(z)$ is the exponential distribution. The potential barrier height reads $\varepsilon = 40$ and temperature is fixed at $D_T = 0.01$. Upper panel corresponds to the symmetric potential $U(x)$ whereas in bottom one it is asymmetric $U_+(x)$.}
	\label{fig:S5}
\end{figure}

\subsection{Other statistics of amplitudes $z_i$}  
In Fig. \ref{fig:S5} we present the average velocity $\langle v \rangle$ versus the spiking rate $\lambda$ of active fluctuations $\eta(t)$ depicted for different amplitude distributions and fixed $\langle \eta(t) \rangle = v_0 = \lambda \zeta = 1$. We compare results obtained for active noise with the skew-normal statistics $\rho(z)$ to the case of Gaussian density $\rho_G(z)$ and the exponential distribution $\rho_e(z)$, namely
\begin{equation}
	\rho_G(z) = \frac{1}{\sqrt{2 \pi \sigma^2}} e^{-(z - \zeta)^2/2 \sigma^2}, \quad \rho_e(z) = \theta(z) \zeta^{-1} e^{-z/\zeta}
\end{equation}
where $\theta(z)$ is the Heaviside step function. $\rho_G(z)$ is the symmetric distribution for which skewness reads $\chi = 0$, whereas the exponential statistics $\rho_e(z)$ with $\chi = 2$ is in fact even more asymmetric than the skew-normal density $\rho(z)$ with $\chi \in (-1,1)$. In the upper panel of Fig. \ref{fig:S5} we present the average velocity $\langle v \rangle$ as a function of spiking rate $\lambda$ for the symmetric potential $U(x)$. In the bottom one the same characteristics is depicted but for the asymmetric (ratchet) substrate $U_+(x)$. Comparison of these two plots illustrates that potential asymmetry can additionally boost the free particle transport. This result is in agreement with the insights presented in Fig. 3.

We can observe that the enormous boost for the free particle transport induced by the periodic potential detected for $\lambda =\lambda_{max} \sim \varepsilon$ is not rooted solely in the asymmetry (non-zero skewness) of active noise amplitude distribution since in both panels this effect is not present for the exponential statistics $\rho_e(z)$. In particular, the particle velocity $\langle v \rangle$ tends either to unity when $\lambda \to 0$ or vanishes if $\lambda \to \infty$. Moreover, when the periodic potential is asymmetric, c.f. the bottom panel with $U_+(x)$, the transport enhancement $\langle v \rangle > v_0$ is discovered even for active noise with amplitudes distributed according to the symmetric Gaussian density $\rho_G(z)$. On the other hand, if the potential is symmetric, see the upper plot with $U(x)$, only active fluctuations with amplitudes drawn from the skew-normal distribution $\rho(z)$ result in the velocity amplification $\langle v \rangle > v_0$. The question about general constraints on the probability density for $\delta$-spikes amplitudes which allow for the emergence of the studied effect is complex and lies beyond the scope of the current paper. Here we only point out important difference between the exponential and skew-normal statistics. In the former one all moments disappear when the mean vanishes $\zeta \to 0$ whereas in the latter mean $\zeta$, variance $\sigma^2$ and skewness are independent parameters. It means that even though $\zeta \to 0$, the variance and skewness can still be finite and fixed.

Our results reveal that periodic potentials can boost the free particle transport driven by white Poissonian noise provided that (i) its amplitude distribution allows for both positive and negative $\delta$-spikes and (ii) the amplitude statistics possesses non-zero skewness or the reflection symmetry of the potential is broken. These are non-trivial conditions for the occurrence of this effect. The fact that active noise with fixed mean and both positive and negative amplitudes can induce significantly greater directed transport than for active noise of equal average but with only positive $\delta$-spikes is in addition highly counter-intuitive. 

\section{Discussion}
The proposed strategy for giant boost of free transport allows to understand why sometimes the existence of periodic structures is beneficial for the transport efficiency. Many transport processes in biological cells are driven by diffusion and consequently their effectiveness is low. Therefore cells have developed another mechanism of movement via microtubules that are asymmetric periodic structures which provide platforms for intracellular transport mediated by molecular motors. These platforms can be formed rapidly in response to cellular needs. They have a half-life of 5-10 minutes \cite{motor} and typically are nucleated and organized by microtubule-organizing centres. The polarized structure of microtubules provide the navigational information necessary to direct cargo to the proper destination in the cell. The biological motor such as a conventional kinesin moves along a microtubule of period 8 nm with a directed velocity 1800 nm/s. Our conclusion is that as Nature teaches us, transport within periodic structures can be many orders of magnitude more efficient than without it. Whether Nature takes advantage of this possibility is another matter, of course.  
 
In summary, we demonstrated a paradoxical effect in which velocity of a free Brownian particle exposed to active fluctuations might be enormously accelerated when it is placed in a periodic potential. Its origin lies in versatile nature of \emph{active nonequilibrium fluctuations} which allow them to fine-tune to the given substrate and transport the particle across the potential barrier where it is able to effortlessly exploit its steepness. This phenomenon should be contrasted with celebrated giant diffusion effect \cite{reimann2001,lindner2016,berezhkovskii2019,spiechowicz2021pre2} in which \emph{thermal equilibrium fluctuations} cooperate with a tilted periodic potential to accelerate diffusion of a particle by many orders of magnitude as compared to free thermal diffusion. 

We considered a paradigmatic model of nonequilibrium statistical physics which comprises numerous realizations \cite{risken} and therefore we expect that our results will inspire a vibrant follow up works. Moreover, the findings may be corroborated experimentally in dissipative optical environment in which the potential barrier can be easily tuned \cite{park2020,paneru2021}. Our results carry impactful consequences not only for microscopic physical systems, but also biological ones such as molecular motors which are \emph{in situ} immersed in unavoidable sea of thermal and active fluctuations. Finally, they may open new avenues for designing ultrafast and efficient micro and nanoscale machines.
\section*{Acknowledgment}
This work has been supported by the Grant NCN No. 2022/45/B/ST3/02619 (J. S.)

\begin{figure}[t]
    \centering
    \includegraphics[width=0.9\linewidth]{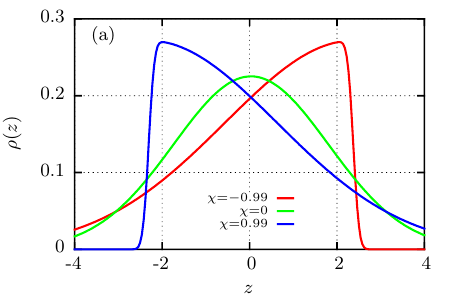}
	\includegraphics[width=0.9\linewidth]{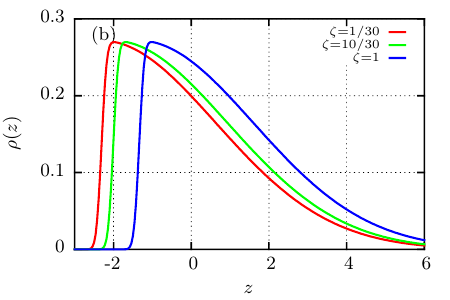}
    \caption{The probability distribution $\rho(z)$ of amplitudes $z_i$ of active nonequilibrium fluctuations $\eta(t)$ is presented in panel (a) for fixed mean $\zeta=1/30$, variance $\sigma^2 = 3.1$  and  different values of skewness $\chi$,  and in panel (b) for various mean $\zeta$ with skewness $\chi=0.99$.}
    \label{fig:S1}
\end{figure}
\appendix
\section{Dimensionless units}
Transforming the equation describing the model into its dimensionless form allows to simplify the analysis because it can reduce the number of parameters appearing before such a procedure. Moreover, the resulting representation is independent of a specific experimental setup allowing to choose the best platform for corroborating theoretical predictions. We start with the Langevin equation for an overdamped Brownian particle in a periodic potential $U(x)$ immersed in both active $\eta(t)$ and thermal $\xi(t)$ bath
\begin{equation}
	\Gamma \dot{x} = - E\,U'(x) + \eta(t) + \sqrt{2 \Gamma k_B T}\, \xi(t)
    \label{seq_1}
\end{equation}
here $\Gamma$ is the friction coefficient, $E$ half of the potential barrier height, $k_B$ the Boltzmann constant and $T$ denotes thermostat temperature. The potential is assumed in the periodic form
\begin{equation}
	U(x) = U(x + L)
\end{equation}
We rescale the position and time in the following way
\begin{align}
	 \hat{x} &= 2\pi \frac{x}{L}, \quad \hat{t}=\frac{t}{\tau_0}, \nonumber \\ \tau_0 &= \frac{L}{100 v_{D_0}} = \frac{L}{100 D_0/L} = \frac{L^2}{100 k_B T/\Gamma}
\end{align}
where $v_{D_0}$ is the characteristic velocity corresponding to free thermal diffusion $D_0 = k_B T/\Gamma$. The additional multiplier $100$ is introduced in the denominator due to technical reasons outlined below. Under such a choice of the scales Eq. (\ref{seq_1}) becomes
\begin{equation}
	\dot{\hat{x}} =-\varepsilon \hat{U}'(\hat{x}) + \hat{\eta}(\hat{t}) + \sqrt{2D_T}\,\hat{\xi}(\hat{t}).
	\label{dimless_model}
\end{equation}
where the dimensionless barrier height is \mbox{$\varepsilon = E/(100 k_B T)$}. The rescaled potential reads
\begin{equation}
	\hat{U}(\hat{x}) = U\left( \frac{L}{2\pi} \hat{x} \right).
\end{equation}
E.g. if $U(x) = \sin{(2\pi x/L)}$ then $\hat{U}(\hat{x}) = \sin{\hat{x}}$ possesses the spatial period $2\pi$. The scaling procedure allows to reduce a number of free parameters by
\begin{equation}
	\gamma = 1, \quad D_T = 0.01.
\end{equation}
Dimensionless thermal noise 
\begin{equation}
	\hat{\xi}(\hat{t}) = \frac{L}{2\pi} \frac{1}{100 k_B T}\, \xi(\tau_0 \hat{t}).
\end{equation}
exhibits the same statistical properties as the dimensional one, i.e., it has the same vanishing mean and correlation function. The rescaled Poissonian shot noise becomes
\begin{equation}
	\hat{\eta}(\hat{t}) = \frac{L}{2\pi} \frac{1}{100 k_B T}\, \eta(\tau_0 \hat{t}),
\end{equation}
and is characterized by the dimensionless spiking rate
\begin{equation}
	\hat{\lambda} = \tau_0 \lambda.
\end{equation}

The error of numerical integration scheme to solve stochastic differential equations driven by Poissonian white shot noise as well as the computation time increases significantly when the frequency $\hat{\lambda}$ of $\delta$-kicks grows \cite{kim2007}. To reduce this drawback we introduced the additional multiplier $100$ in the definition of characteristic time scale $\tau_0$ which allows to limit the range of spiking rate $\hat{\lambda}$ needed in this study. As in the main text only the dimensionless quantities are used we omit there the hat notation.
\begin{figure}[t]
    \centering
    \includegraphics[width=0.9\linewidth]{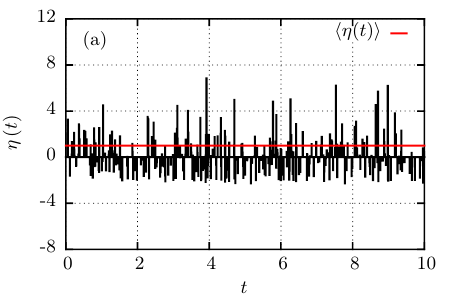}\\
    \includegraphics[width=0.9\linewidth]{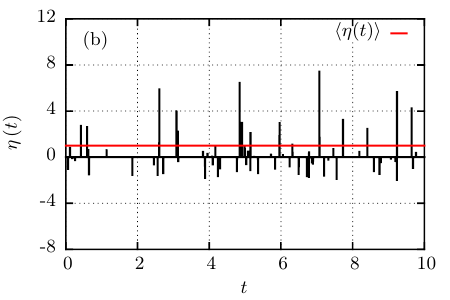}\\
    \includegraphics[width=0.9\linewidth]{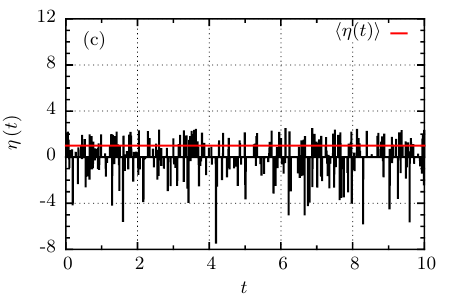} 
    \caption{Three exemplary realizations of active nonequilibrium noise $\eta(t)$ for different parameters: (a) $\lambda=30$, $\zeta=1/30$, $\sigma^2=3.1$ and $\chi=0.99$; (b) $\lambda=6$, $\zeta=1/6$, $\sigma^2=3.1$ and $\chi=0.99$; (c) $\lambda=30$, $\zeta=1/30$, $\sigma^2=3.1$ and $\chi=-0.99$. Solid red line corresponds to the average $\langle \eta(t) \rangle = 1$.}
    \label{fig:S2}
\end{figure}

\section{Parametrization of amplitude distribution $\rho(z)$}
The skew-normal distribution \cite{azz} is usually defined in terms of a location $\mu$ representing the shift, a scale $\omega$ proportional to the variance and a parameter $\alpha$ describing the shape. The probability density function then reads
\begin{equation}
    \rho(z) = \frac{2}{\sqrt{2\pi \omega^2}}e^{-\frac{(z-\mu)^2}{2\omega^2}} \int_{-\infty}^{\alpha[(z-\mu)/\omega]} ds \, \frac{1}{2\pi}e^{-\frac{s^2}{2}}.
\end{equation}
The quantities $\mu$, $\omega$ and $\alpha$ can be represented by more intuitive parameters, namely, the mean $\mu \to \zeta = \langle z_i \rangle$, variance 
$\omega \to \sigma^2 = \langle (z_i - \langle z_i \rangle)^2 \rangle$ and skewness $\alpha \to \chi = \langle [(z_i - \langle z_i \rangle)/\sigma]^3 \rangle$ of the distribution $\rho(z)$. The average amplitude $\zeta$ allows for direct control of mean bias $\langle \eta(t) \rangle = v_0 = \lambda \zeta$ of nonequilibrium noise, the variance $\sigma^2$ describes the mean square deviation of $\delta$-spikes and skewness $\chi$ measures the asymmetry of skew-normal distribution. Expressions for the location $\mu$, scale $\omega$ and shape $\alpha$ in terms of these parameters are as follows \cite{generacja,generacja2}
\begin{subequations}
\begin{align}
\begin{split}
\alpha&=\frac{\delta}{\sqrt{1-\delta^2}}, \\
\end{split}\\
\begin{split}
\omega&=\sqrt{\frac{\sigma^2}{1- 2\delta^2/\pi}}, \\
\end{split}\\
\begin{split}
\mu&=\zeta-\delta\sqrt{\frac{2\sigma^2}{\pi(1-2\delta^2/\pi)}},\\
\end{split}
\end{align}
\label{eq_S_def}
\end{subequations}
where $\delta$ is defined as
\begin{equation}
    \delta=\text{sgn}(\chi)\sqrt{\frac{|\chi|^{2/3}}{(2/\pi)\{[(4-\pi)/2]^{2/3}+|\chi|^{2/3}\}}}.
    \label{eq_S_delta}
\end{equation}
In Fig. \ref{fig:S1} we present the probability density function $\rho(z)$ for amplitudes of active nonequilibrium noise $\eta(t)$ as a function of its parameters, i.e. mean $\zeta$, variance $\sigma^2$ and skewness $\chi$. In panel (a) the impact of skewness $\chi$ for the fixed mean $\zeta = 1/30$ is shown whereas in (b) the influence of mean $\zeta$ is illustrated for $\chi = 0.99$.

\section{Realizations of active fluctuations $\eta(t)$}
In Fig. \ref{fig:S2} we demonstrate three exemplary realizations of active nonequilibrium noise $\eta(t)$ for different values of its parameters. For all presented cases the statistical bias is fixed to $\langle \eta(t) \rangle = v_0 = 1$. Panel (a) corresponds to the regime of optimal transport $\langle v \rangle$ for the potential barrier height $\varepsilon = 40$, see Fig. 2 in the main text, the spiking rate $\lambda = 30$, mean amplitude $\zeta = 1/30$, variance $\sigma^2 = 3.1$ and skewness $\chi = 0.99$. In panel (b) the frequency is five times smaller $\lambda = 6$ and therefore $\zeta = 1/\lambda = 1/6$ to satisfy the condition $\langle \eta(t) \rangle = \lambda \zeta = v_0 = 1$. The reader can indeed infer that the parameter $\lambda$ describes the frequency of $\delta$-spikes whereas an increase of $\zeta$ has two-fold effect: (i) the positive amplitudes $z_i > 0$ grows and (ii) the negative ones $z_i < 0$ are reduced. Finally, in panel (c) we present the impact of skewness inversion $\chi \to -\chi$. As it is visualized in this plot such operation corresponds to reflection of the realization about the axis $\eta(t) = \zeta$.

\end{document}